%% file: 6v2.tex
\def\answ{b }
\input harvmac
\input mssymb

\input labeldefs.tmp
\writedefs
\overfullrule=0pt

% redefine figures ... 
\input epsf
\def\fig#1#2#3{
\xdef#1{\the\figno}
\writedef{#1\leftbracket \the\figno}
\nobreak
\par\begingroup\parindent=0pt\leftskip=1cm\rightskip=1cm\parindent=0pt
\baselineskip=11pt
\midinsert
\centerline{#3}
\vskip 12pt
{\bf Fig. \the\figno:} #2\par
\endinsert\endgroup\par
\goodbreak
\global\advance\figno by1
}
\newwrite\tfile\global\newcount\tabno \global\tabno=1
\def\tab#1#2#3{
\xdef#1{\the\tabno}
\writedef{#1\leftbracket \the\tabno}
\nobreak
\par\begingroup\parindent=0pt\leftskip=1cm\rightskip=1cm\parindent=0pt
\baselineskip=11pt
\midinsert
\centerline{#3}
\vskip 12pt
{\bf Tab. \the\tabno:} #2\par
\endinsert\endgroup\par
\goodbreak
\global\advance\tabno by1
}
\def\der{\partial}
\def\d{{\rm d}}
\def\e#1{{\rm e}^{#1}}
\def\E#1{{\rm e}^{\textstyle #1}}%
\def\cn{\mathop{\rm cn}\nolimits}
\def\dn{\mathop{\rm dn}\nolimits}
\def\sn{\mathop{\rm sn}\nolimits}
%\font\rfont=cmss8

%

%
%%%%%%%%%%%%%%%%%%%%%%%
% preprint # in refs ?
\def\pre#1{ (preprint {\tt #1})}%use this to give preprint # in refs
%\def\pre#1{}%use this NOT to give preprint # in refs
%%%%%%%%%%%%%%%%%%%%%%%%%%%%%%%%%%%%%%%%%%%%%%%%%%%%%%%%%%%%
%
%References
%
%\lref\XXX{N.N. XXX, {\it title }, .}
% \refs{\..{--}\...}
%
\lref\ICK{A.G.~Izergin, D.A.~Coker and V.E.~Korepin,
{\it J. Phys.} A 25 (1992), 4315.}
\lref\Ize{A.G.~Izergin, {\it Sov. Phys. Dokl.} 32 (1987), 878.}
\lref\Suth{ B. Sutherland, {\it PRL} 19 (1967), 103.}
\lref\Kor{V.E. Korepin, {\it Commun. Math. Phys} 86 (1982), 391.}
\lref\Bax{R.J.~Baxter, {\sl Exactly Solved Models in
Statistical Mechanics} (San Diego, CA: Academic).}
\lref\Kup{G.~Kuperberg, 
{\it Internat. Math. Res. Notices} 3 (1996), 139\pre{math.CO/9712207}.}
\lref\Sog{K.~Sogo, {\it Journal of the Physical Society of Japan}
62, 6 (1993), 1887.}
\lref\Hir{R. Hirota {\it Journ Phys. Soc. Japan} 56 (1987), 4285.}
\lref\Wieg {O. Lipan, P.B. Wiegmann, A. Zabrodin,  preprint {\tt solv-int/9704015}.}
\lref\WZ {P.B. Wiegmann, A. Zabrodin, preprint {\tt hep-th/9909147}.}
\lref\WZK{I. Krichever, O. Lipan, P. Wiegmann, A. Zabrodin, preprint {\tt hep-th/9604080}.}
\lref\UT{ K. Ueno, K.Takasaki {\it Adv. Studies in Pure Math.} 4 (1984), 1.}
\lref\AvM{M.~Adler and P.~van~Moerbeke, {\it Duke Math. Journal}
80 (1995), 863\pre{solv-int/9706010};
preprint {\tt math.CO/9912143}.}
\lref\To{M. Toda {\it Journ. Phys. Soc. Japan} 22 (1967), 431;
{\it Prog. Theor. Phys. Suppl.} 45 (1970), 174.}
\lref\Fla{H. Flaschka {\it Phys. Rev.} B 9 (1974); {\it Prog. Theor.
Phys.} 51 (1974), 703.}
\lref\GMMMO{A.~Gerasimov, A.~Marshakov, A.~Mironov, A.~Morozov
and A.~Orlov, {\it Nucl. Phys.} B 357 (1991), 565.}
\lref\L{E.~Lieb, {\it Phys. Rev. Let. } 18 (1967), 692.}
\lref\Li{E.~Lieb, {\it Phys. Rev. Let.} 18 (1967) 1046.}
\lref\Lie{E.~Lieb, {\it Phys. Rev. Let.} 19 (1967) 108.}
\lref\Lieb{E.~Lieb,  {\it  Phys. Rev.} 162 (1967) 162.}
\lref\Kas{P.W.~Kasteleyn, {\it Physica} 27 (1960), 1209.}
\lref\Fi{M.E.~Fisher, {\it Phys. Rev.} 124 (1961), 1664.}
\lref\LW{W.T.~Lu and F.Y.~Wu, {\it Phys. lett. A } 259 (1999), 108.}
\lref\JPS{W.~Jockush, J.~Propp and P.~Shor, preprint {\tt math.CO/9801068}.}
% arctic circle theorem. note paper is 95, what is journal ref?
%
\lref\CEP{H.~Cohn, N.~Elkies and J.~Propp, {\it Duke Math. Journal}
85 (1996), 117.}
% arctic circle and local probabilities => heterogeneity inside circle
%
\lref\EKLP{N.~Elkies, G.~Kuperberg, M.~Larsen and J.~Propp,
{\it Journal of Algebraic Combinatorics} 1 (1922), 111; 219.}
% ASM & dominos
%
\lref\Br{D.M.~Bressoud, {\sl Proofs and Confirmations:
The Story of the Alternating Sign Matrix Conjecture},
Cambridge University Press, Cambridge, 1999}
\lref\BP{D.~Bressoud and J.~Propp, {\it Notices of the AMS} June/July (1999), 637.}
\lref\Deift{P.~Deift, T.~Kriecherbauer, K.T-R.~McLaughlin, S.~Venakides and 
X.~Zhou, {\it Commun. on Pure and Applied Math.} 52
(1999), 1491.}
\lref\BDE{G.~Bonnet, F.~David and B.~Eynard, preprint {\tt cond-mat/0003324}.}
% multi-cut matrix models
\lref\KZJ{V.~Korepin and P.~Zinn-Justin, preprint {\tt cond-mat/0004250}.}
\lref\BKaz{E.~Br\'ezin and V.~Kazakov, preprint {\tt math-ph/9909009}.}
\lref\KazZJ{V.~Kazakov and P.~Zinn-Justin,
{\it Nucl. Phys.} B546 (1999), 647\pre{hep-th/9808043}.}
\lref\BDJ{J.~Baik, P.~Deift and K.~Johansson,
{\it J. Amer. Math. Soc.} 12 (1999) no. 4, 1119\pre{math.CO/9810105}.}
\lref\Ver{A.M.~Vershik and S.V.~Kerov,
{\it Soviet. Math. Dokl.} 18 (1977), 527.}
% first paper on Plancherel measure
%
\lref\Joh{K.~Johansson, preprint {\tt math.CO/9906120}.}
\lref\BOO{A.~Borodin, A.~Okounkov and G.~Olshanski,
preprint {\tt math.CO/9905032}.}
\lref\DK{M.R.~Douglas and V.A.~Kazakov,
{\it Phys. Lett.} B319 (1993), 219.}
\lref\KKSW{V.A.~Kazakov, M.~Staudacher and T.~Wynter,
{\it Commun. Math. Phys.} 177 (1996), 451; 179 (1996), 235;
{\it Nucl. Phys.} B471 (1996), 309\semi
I.~Kostov, M.~Staudacher and T.~Wynter,
{\it Commun. Math. Phys.} 191 (1998), 283.}
\lref\Ken{R.~Kenyon, {\sl The planar dimer model with boundary:
a survey}, preprint\hfil\break
({\tt http://topo.math.u-psud.fr/$\sim$kenyon/papers.html}).}
% review on effects of BC on domino tilings
%%%%%%%%%%%%%%%%%%%%%%%%%%%%%%%%%%%%%%%%%%%%%%%%%%%%%%%%%%%%%%%%%%%%%
\Title{
\vbox{\baselineskip12pt\hbox{YITP-00-21}
\hbox{{\tt math-ph/0005008}}}
}
{{\vbox {
\vskip-10mm
\centerline{\bf Six-Vertex Model with Domain Wall Boundary Conditions}
\vskip2pt
\centerline{\bf and One-Matrix Model}
}}}
\medskip
\centerline{P.~Zinn-Justin\footnote{*}{e-mail: 
{\tt pzinn@insti.physics.sunysb.edu}}}\medskip
\centerline{\sl C.N.~Yang Institute for Theoretical Physics}
\centerline{\sl State University of New York at Stony Brook}
\centerline{\sl Stony Brook, NY 11794--3840, USA}
\vskip .2in
% abstract
\noindent The partition function of the six-vertex model on a
square lattice
with domain wall boundary conditions (DWBC)
is rewritten as a hermitean one-matrix model or a discretized version of
it (similar to sums over Young diagrams),
depending on the phase. The expression
is exact for finite lattice size, which is equal to the size of the
corresponding matrix. In the thermodynamic limit,
the matrix integral is computed using
traditional matrix model techniques, 
thus providing a complete treatment of the
bulk free energy of the six-vertex model with DWBC in the
different phases.
In particular, in the anti-ferroelectric phase,
the bulk free energy and a subdominant correction
are given exactly in terms of elliptic theta
functions.
\Date{05/2000}
%\draft

\newsec{Introduction}
In \KZJ, V.~Korepin and the author brought up
the issue of the sensitivity of the six-vertex model to its
boundary conditions (even in the thermodynamic limit).
The motivation came mostly from some recent work on domino tilings
\refs{\JPS,\CEP,\Ken}, in which boundary conditions seemed to affect
greatly the typical arrangement of dominos. The problem
of counting domino tilings is equivalent to the six-vertex model
with particular Boltzmann weights; this is schematically described
on Fig.~\domino.
Therefore it seems natural to investigate the corresponding problem
for the general six-vertex model with arbitrary weights.
\fig\domino{Correspondence between vertices
of the six-vertex model and small patches of a domino 
tiling.}{\epsfxsize=8cm\epsfbox{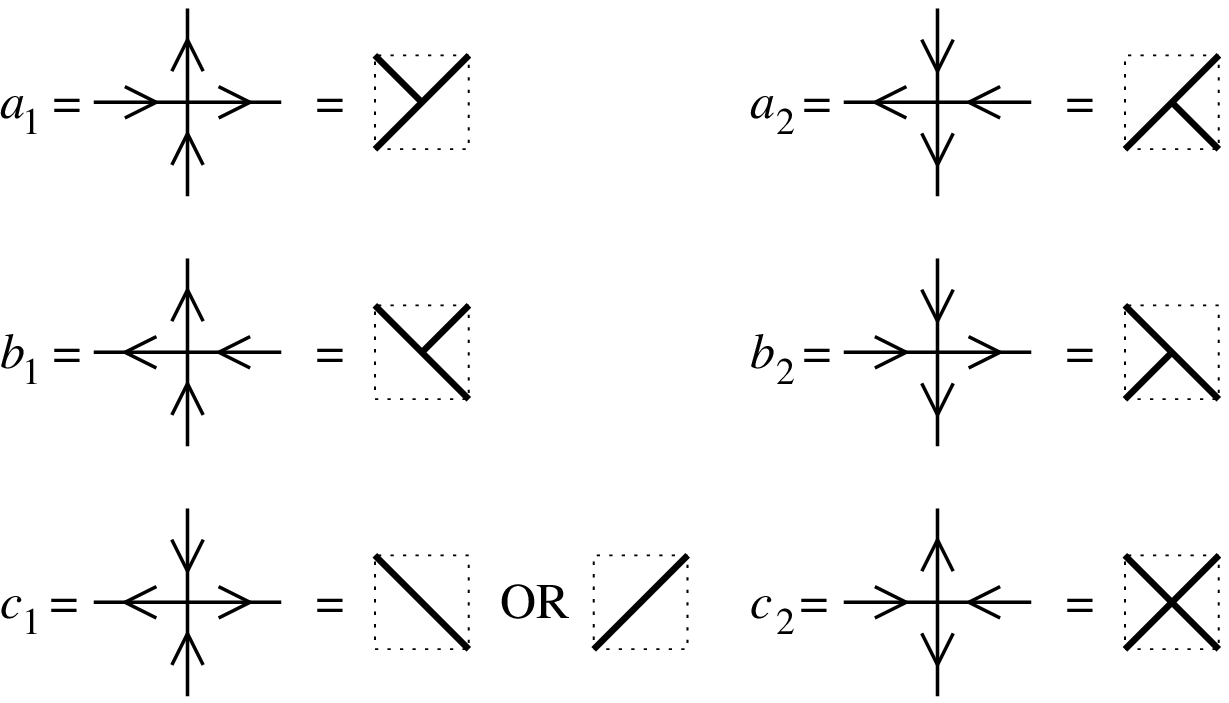}}

The usual studies of the six-vertex model (see \Bax\ and
references therein) are made by assuming periodic boundary conditions
(PBC). In \KZJ, different boundary conditions, the so-called
domain wall boundary conditions (DWBC), were used (Fig.~\dominob a), and the
thermodynamic limit of the model was investigated using determinant
formulae for the partition function \refs{\Ize,\ICK}. The main result
found was an expression for the bulk free energy in the disordered phase
of the model, which is different from
the usual expression for the case of periodic boundary conditions.
It should be noted that the DWBC correspond to the Aztec shape in the
domino tiling language (see Fig.~\dominob), which is precisely
the type of tiling which was considered in \refs{\JPS,\CEP}.
\fig\dominob{a) A configuration of the six-vertex model with DWBC,
and b) one possible corresponding tiling of the Aztec diamond.}
{\epsfxsize=8.5cm\epsfbox{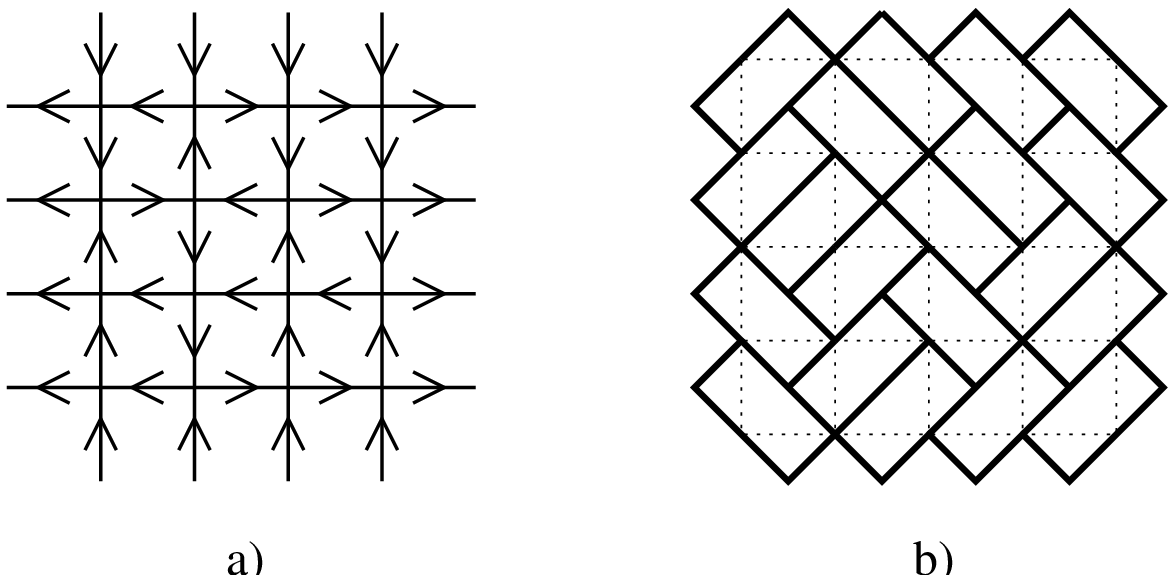}}

Here, we use a new method to compute the bulk free energy
with DWBC in all phases of the model; in particular,
we obtain an independent confirmation of the results of \KZJ.
In section 2, starting from the
determinant formula for the partition function, we shall rewrite
the latter as a matrix integral, but with a measure
on the space of hermitean matrices which is not necessarily smooth. 
In the
disordered phase (section 4), the measure will turn out to be smooth, whereas
in the ferroelectric and anti-ferroelectric phase (section 3 and 5)
it will be discrete (when expressed in terms of the eigenvalues).
The size $N$ of the matrices is the size of the original
square lattice, and therefore the thermodynamic limit can be
investigated using tools from large $N$ matrix models.
Since the results of section 5 (concerning the
anti-ferroelectric phase) are new, they are analyzed in more detail
by considering various limits of the parameters,
and the subleading correction of the free energy is calculated.

\newsec{Properties of the determinant formula}
We use the same notations as in \KZJ. We consider the homogeneous
six-vertex model,
with the following parameterization of the Boltzmann weights
attached to the vertices:
\eqn\wei{
a=\sinh(t-\gamma)\qquad
b=\sinh(t+\gamma)\qquad
c=\sinh(2\gamma)
}
The domain wall boundary conditions (DWBC) mean that 
external horizontal
arrows are outgoing, whereas external vertical arrows are
incoming (Fig.~\dominob a). These boundary conditions only
exist for square lattices.
In \refs{\Ize,\ICK}, it was shown that the partition function of the
six-vertex model with DWBC on a $N\times N$ lattice could be written as:
\eqn\detrep{
Z_N={(\sinh(t+\gamma)\sinh(t-\gamma))^{N^2}
\over\left(\prod_{n=0}^{N-1} n!\right)^2} \tau_N
}
where $\tau_N$ is a H\"ankel determinant:
\eqn\deftau{
\tau_N=
\det_{1\le i,k\le N}
\left[{\d^{i+k-2}\over\d t^{i+k-2}}\phi(t)\right]
}
Here,
\eqn\defphi{
\phi(t)\equiv{\sinh(2\gamma)\over\sinh(t+\gamma)\sinh(t-\gamma)}
}
It is known that such determinants are tau-functions
of the Toda semi-infinite
chain hierarchy in terms of appropriate parameters. Here,
as a function of $t$, the $\tau_N$ satisfy the ususal Toda
equations under the bilinear form \refs{\Sog,\KZJ}:
\eqn\hirota{
\tau_N\tau_N''-\tau_N'{}^2=\tau_{N+1}\tau_{N-1}\qquad \forall N\ge 1
}
This equation was used in \KZJ\ to derive the bulk free energy of the model
in the ferroelectric and disordered phase by making an appropriate
Ansatz on the large $N$ form of $\tau_N$. Unfortunately, the
Ansatz in the anti-ferroelectric phase is not that simple, as we
shall see, and would be hard to justify at this point.

We shall therefore use another approach here, based on the equivalence
of H\"ankel determinants with one-matrix models \refs{\GMMMO,\AvM}.
Let us write formally
$\phi(t)$ as a Laplace transform:
\eqn\lap{
\phi(t)=\int\d m(\lambda) \e{t\lambda}
}
where $\d m(\lambda)$ is a measure. We then notice that the
derivatives of $\phi(t)$ are the moments:
\eqn\mom{
{\d^i\over\d t^i}\phi(t)=\int\d m(\lambda) \lambda^i \e{t\lambda}
}
Inserting this into \deftau\ leads to:
\eqn\dermm{
\tau_N=\int\d m(\lambda_1)\ldots \d m(\lambda_N)
\sum_{\sigma\in{\cal S}_N}(-1)^\sigma \prod_{i=1}^N
\left[\e{t\lambda_i} \lambda_i^{i+\sigma(i)-2}\right]
}
We see that appears naturally the Van der Monde determinant
$\Delta(\lambda_i)=\det(\lambda_i^{j-1})=\prod_{i<j}(\lambda_i-\lambda_j)$.
After a few elementary manipulations we find:
\eqn\dermmb{
\tau_N={1\over N!}\int\d m(\lambda_1)\ldots \d m(\lambda_N)
\Delta(\lambda_i)^2 \e{t\sum_i \lambda_i}
}
If $\d m(\lambda)$ is a smooth positive measure of the
form $\d m(\lambda)=\d\lambda\, \e{-V(\lambda)}$, then we recognize
in \dermmb\ the expression in terms of its eigenvalues
of the matrix integral:
\eqn\dermmc{
\tau_N\sim\int \d M \,\E{\tr [tM+V(M)]}
}
where $M$ is a hermitean $N\times N$ matrix, and $\d M$ is the flat
measure.

As we shall see, if the measure is not smooth, we shall end up with
expressions which can still be treated using appropriately
adapted matrix model techniques. This is typically the case of discrete
measures that appear in sums over Young diagrams 
\refs{\Ver,\DK,\KKSW,\KazZJ,\BDJ,\Joh,\BOO,\BKaz}.

Expressions of the type \dermmb\ have been widely studied in the
literature (on random matrices in particular). One important
goal is to find their large $N$ asymptotic behavior.
Here we shall mention the simplest method to find their leading
large $N$ behavior: the saddle point method. The basic idea
is that $\log\Delta(\lambda_i)^2$, being a sum of $\sim N^2$ terms,
scales as $N^2$ in the large limit, whereas there are only $N$
variables of integration. Therefore the integral is dominated by a
saddle point. An important remark is that, in order
to find the saddle point, we must
write our action (i.e.\ log of the function integrated)
in such a way that all terms are of the same order $N^2$.
Here, the term $t\sum_i \lambda_i$ is na\"\i vely of order $N$, and
we reach the important conclusion that the $\lambda_i$ will scale as
\eqn\scal{
\lambda_i\propto N\mu_i
}
After the change of variables $\lambda_i\to\mu_i$, one can use the
saddle point approximation, which gives us access to the function $f$
defined by
\eqn\deff{
f=\lim_{N\to\infty} {\log(\tau_N/c_N)\over N^2}
}
where $c_N\equiv(\prod_{n=0}^{N^2} n!)^2$. $f$ is essentially
the bulk free energy, cf Eq.~\detrep. Note that the saddle point is
a very crude approximation in the sense that it does
not naturally allow for a systematic computation of subleading
corrections; however it will be sufficient for our purposes.
We now proceed with a separate discussion of the different
phases of the model.

\newsec{Ferroelectric phase}
This is the phase in which the weights are given by \wei\ with
$t$ and $\gamma$ real, $|\gamma|<t$.
We use the following decomponsition:
\eqn\lapc{
\phi(t)={\sinh(2\gamma)\over\sinh(t+\gamma)\sinh(t-\gamma)}
=4\sum_{l=0}^\infty \e{-2tl}\sinh(2\gamma l)
}
We are in the situation where the measure $\d m$ is discrete. The determinant
takes the form
\eqn\discrF{
\tau_N=2^{N^2}\sum_{l_1,\ldots,l_N=0}^\infty \Delta(l_i)^2 \e{-2t\sum_i l_i}
\prod_i \sinh(2\gamma l_i)
}
(we have neglected here, as in all subsequent calculations, constant factors
which manifestly do not contribute to the bulk free energy).
This expression is very close
to what one encounters when studying the Plancherel measure (or
other similar measures) on Young
diagrams \Ver. In the context of Young diagrams, the $l_i$
represent the shifted highest weights $l_i=m_i+N-i$, where the $m_i$
are the usual highest weights (sizes of the rows of the diagram),
and one is usually interested in the {\it limiting shape} of the Young diagram
when its size is sent to infinity. There has
been a lot of work on this type of expressions, both in the mathematical
literature \refs{\Ver,\BDJ,\Joh,\BOO} (the recent work being concerned
with {\it fluctuations} around the limiting shape, which we shall
not discuss here)
and the physical literature \refs{\DK,\KKSW,\KazZJ,\BKaz}.
One relevant observation from \DK\ is the following:
after the rescaling $\mu=l/N$, all sums look like Rieman sums and
one is tempted to replace them with integrals, and then apply the
saddle point method. This is correct {\it on condition} that
one imposes an additional constraint coming from the discreteness of the $l_i$.
In Eq.~\discrF, all $l_i$ must be distinct integers (due to the
Van der Monde determinant), and therefore
\eqn\ineq{
|l_i-l_j|\ge 1\qquad\forall i\ne j
}
If we introduce the density $\rho(\mu)\,\d\mu$ of the $\mu_i=l_i/N$, normalized so
that $\int\rho(\mu)\,\d\mu=1$, then \ineq\ implies that it
must satisfy the inequality
\eqn\ineqb{
\rho(\mu)\le 1
}
In general, when the $l_i$ are trapped in a well of the potential (as is the
case here), there will be a saturated region at the bottom of the well
where $\rho(\mu)=1$, and an unsaturated region where $\rho(\mu)<1$.

Let us now proceed with the solution. Once the rescaling $\mu_i=l_i/N$
is performed, one notices that up to corrections exponentionally
small in $N$, $\sinh(2\gamma N \mu_i)\approx {1\over2} \e{2|\gamma| N\mu_i}$.
Therefore
\eqn\planch{
\tau_N\approx c'_N\, 2^{N^2}\sum_{\mu_1,\ldots,\mu_N\in{1\over N}{\Bbb Z}_+}
\Delta(\mu_i)^2 \e{-2N(t-|\gamma|)\sum_i \mu_i}
}
where $c'_N\equiv N^{N^2}$.
Of course, once this simplification is made, we regognize a well-known
expression; in fact, going back now to the original variables $l_i$ one
can compute $\tau_N$ directly using the Cauchy identity for 
Schur functions. However, to emphasize the similarity with the other
phases (which do not possess such a simple group-theoretic interpretation),
we shall use the saddle point method, following
the solution of \BKaz.
Since $\tau_N$ only depends on $t-|\gamma|$, we temporarily
set $\gamma=0$.

\def\a{\alpha}\def\b{\beta}
The support of the saddle point density $\rho(\mu)$ is expected to be of the form
$[0,\b]$; the saturated region is $[0,\a]$, whereas the unsaturated region
is $[\a,\b]$. We define the resolvent
\eqn\defres{
\omega(z)=\int_0^\b {\d\mu \,\rho(\mu)\over z-\mu}
}
for all complex $z\not\in[0,\b]$. The saddle point equations can be
written in terms of $\omega$:
\eqn\speF{
\omega(\mu+i0)+\omega(\mu-i0)=2t\qquad\forall \mu\in[\a,\b]
}
In order to solve the equation, we first remove the logarithmic
cut of $\omega$ with the redefinition: $\tilde{\omega}(z)=\omega(z)-
\log{\mu\over\mu-\a}$. $\tilde{\omega}(z)$ is analytic everywhere
except on $[\a,\b]$ and satisfies
\eqn\speFb{
\tilde{\omega}(\mu+i0)+\tilde{\omega}(\mu-i0)=2t-2\log{\mu\over\mu-\a}
}
This completely determines it to be:
\eqn\solspeF{
\tilde{\omega}(z)=t-\sqrt{(z-\a)(z-\b)}\int_{\a-i0}^{\b-i0}
{\d z'\over 2i\pi(z-z')\sqrt{(z'-\a)(z'-\b)}} \log{z'\over z'-\a}
}
After some calculations, we find that
\eqn\solspeFb{
\omega(z)=t-2\log\left[{\sqrt{\b(z-\a)}+\sqrt{\a(z-\b)}\over\sqrt{z(\b-\a)}}\right]
}
The endpoints $\a$ and $\b$ are determined by imposing $\omega(z)\sim {1\over z}$
as $z\to\infty$. This gives rise to two equations:
\eqn\bcspeF{
\cases{
t=\log{\sqrt{\b}+\sqrt{\a}\over\sqrt{\b}-\sqrt{\a}}\cr
\sqrt{\a\b}=1\cr
}
}
whose solution is:
\eqn\bcspeFb{
\a= \coth {t\over 2}\qquad \b=\tanh {t\over 2}
}
In order to conclude, one expands further the function $\omega(z)$:
\eqn\expomF{
\omega(z)={1\over z}+{\a+\b\over4}{1\over z^2}+\cdots
}
and uses the fact that
\eqn\corF{
{\der f\over\der t}=-2 \left< \mu \right> = -{\a+\b\over 2}=\coth t
}
Integrating once and restoring $\gamma$,
we have the final result
\eqn\finalF{
\e{f}={1\over\sinh(t-|\gamma|)}
}
which coincides with what was found in \KZJ.

\newsec{Disordered phase}
In this phase, one usually rewrites the weights
\eqn\weiD{
a=\sin(\gamma-t)\qquad
b=\sin(\gamma+t)\qquad
c=\sin(2\gamma)
}
with redefined parameters $t$ and $\gamma$, $|t|<\gamma$, and the function 
$\phi(t)=\sin(2\gamma)/(\sin(t-\gamma)\sin(t+\gamma))$; the partition
function is then given by
\eqn\detrepD{
Z_N={(\sin(\gamma+t)\sin(\gamma-t))^{N^2}
\over\left(\prod_{n=0}^{N-1} n!\right)^2} \tau_N
}
with $\tau_N$ still given by \deftau.
The Laplace transform is:
\eqn\lapb{
\phi(t)={\sin(2\gamma)\over\sin(\gamma+t)\sin(\gamma-t)}
=\int_{-\infty}^{+\infty}\d\lambda\,\e{t\lambda} {\sinh {\lambda\over2}
(\pi-2\gamma)\over\sinh {\lambda\over2}\pi}
}
This time the measure is smooth and $\tau_N$ is a matrix integral
in the usual sense.

We must now rescale the variables $\lambda_i$. We choose
to define $\mu_i=\gamma\lambda_i/N$. Then:
\eqn\dis{
\tau_N= c'_N\,\gamma^{-N^2}
\int_{-\infty}^{+\infty}\d\mu_1\ldots\d\mu_N \Delta(\mu_i)^2
\prod_{i=1}^N\left[
{\sinh N\mu_i
({\pi\over2\gamma}-1)\over\sinh N\mu_i{\pi\over 2\gamma}}
\e{N {t\over\gamma}\mu_i}\right]
}
One then simplifies the potential by using:
${\sinh N\mu
({\pi\over2\gamma}-1)\over\sinh N\mu{\pi\over 2\gamma}}
\sim \e{-N|\mu|}$. Therefore,
\eqn\disb{
\tau_N\approx c'_N\, \gamma^{-N^2}
\int_{-\infty}^{+\infty}\d\mu_1\ldots\d\mu_N \Delta(\mu_i)^2
\e{N\sum_i ({t\over\gamma}\mu_i-|\mu_i|)}
}
Note that the matrix integral only depends on the ratio $\zeta\equiv 
t/\gamma$.

The matrix model \disb\ is fairly simple and can be solved
easily in the large $N$ limit via the saddle point method.
One introduces again the saddle point density of eigenvalues $\rho(\mu)\, \d\mu$,
normalized so that $\int\rho(\mu)\,\d\mu=1$. The support of $\rho(\mu)$
is assumed to be a single interval $[\a,\b]$ ($\a<0<\b$),
due to the shape of the potential (single well centered around $0$).
The resolvent is defined as before. The saddle point equations read:
\eqn\speD{
\omega(\mu+i0)+\omega(\mu-i0)=-\zeta+{\rm sign}(\mu)\qquad\forall
\mu\in[\a,\b]
}
where the right hand side is simply the derivative of the potential.
The solution of this equation:
\eqn\solspeD{
\omega(z)=
{1-\zeta\over2}+{2\over i\pi}\log
\left[\sqrt{\b(z-\a)}-i\sqrt{-\a(z-\b)}\over\sqrt{z(\b-\a)}\right]
%{1\over i\pi}\log
%\left[{(z+i\sqrt{-ab}+\sqrt{(z-a)(z-b)})^2\over z(\sqrt{z-a}+\sqrt{z-b})^2}
%{\sqrt{b}+i\sqrt{-a}\over\sqrt{b}-i\sqrt{-a}}\right]-{1-\zeta\over2}
}
is very similar to the ferroelectric phase;
and the rest of the calculation goes along the same lines.

Requiring that $\omega(z)\sim{1\over z}$ as $z\to\infty$,
we obtain the $2$ equations:
\eqn\bcspeD{
\cases{
1-\zeta={2\over i\pi}\log{\sqrt{\b}+i\sqrt{-\a}\over\sqrt{\b}-i\sqrt{-\a}}\cr
\sqrt{-\a\b}=\pi\cr
}
}
which we solve for $\a$ and $\b$:
\eqn\bcspeDb{
\a=-\pi \tan {\pi\over4}(1-\zeta)\qquad
\b=\pi\tan{\pi\over 4}(1+\zeta)
}
Noting that
\eqn\cor{
{\der f\over\der\zeta}=\left<{1\over N}\tr M\right>={\a+\b\over 4}
}
we find
\eqn\final{
f=-\log\cos{\pi\over2}\zeta+{\rm cst}
}
We shall not discuss how to fix the constant of integration, since
this will be addressed in the next section in a more general setting.
Reintroducing the $\gamma$ dependence coming from Eq.~\disb,
we have the final expression:
\eqn\finalD{
\e{f}={\pi\over 2\gamma}{1\over\cos{\pi t\over 2\gamma}}
}
which reproduces the result of \KZJ.

\newsec{Anti-ferroelectric phase}
We finally study the most interesting phase, in which the
weights are given by
\eqn\weiAF{
a=\sinh(\gamma-t)\qquad
b=\sinh(\gamma+t)\qquad
c=\sinh(2\gamma)
}
with $|t|<\gamma$, and the partition function by
\eqn\detrepAF{
Z_N={(\sinh(\gamma+t)\sinh(\gamma-t))^{N^2}
\over\left(\prod_{n=0}^{N-1} n!\right)^2} \tau_N
}
with $\phi(t)=\sinh(2\gamma)/
(\sinh(\gamma+t)\sinh(\gamma-t))$.

\subsec{Bulk free energy}
We have the expansion
\eqn\lapc{
\phi(t)={\sinh(2\gamma)\over\sinh(\gamma+t)\sinh(\gamma-t)}
=2\sum_{l=-\infty}^{+\infty} \e{2tl}\e{-2\gamma |l|}
}
We perform the rescaling $\mu_i=2\gamma l_i/N$ and find that
$\tau_N$ takes the form:
\eqn\discrAF{
\tau_N=c'_N\,\gamma^{-N^2}
\sum_{\mu_1,\ldots,\mu_N\in{2\gamma\over N}{\Bbb Z}} 
\Delta(\mu_i)^2 
\e{N\sum_i ({t\over\gamma}\mu_i-|\mu_i|)}
}
The remarkable feature is that Eq.~\discrAF\ is identical
to Eq.~\disb\ up to the discrete nature of the variables!
We shall comment on this later.

The situation is a bit more complicated than in the previous cases,
since we now expect a saturated region $[\a',\b']$ at the bottom of the well
($\a'<0<\b'$)
and {\it two} unsaturated regions $[\a,\a']$ and $[\b',\b]$ on each side.
This is a two-cut situation, which is in fact the reason
why the na\"\i ve approach
of \KZJ\ fails in the anti-ferroelectric phase (see section 5.3 for more
on this).
Let us define as before $\zeta=t/\gamma$, the density $\rho(\mu)$ and
its resolvent $\omega(\mu)$. The constraint coming from the discreteness
of the $\mu_i$ reads
\eqn\constr{
\rho(\mu)\le {1\over 2\gamma} \qquad\forall\mu
}
Therefore we have in the saturated region the equation
\eqn\sat{
\rho(\mu)={1\over 2i\pi}(\omega(\mu-i0)-\omega(\mu+i0))={1\over 2\gamma}
\qquad\forall\mu\in[\a',\b']
}
whereas in the unsaturated regions, the saddle point equations are
\eqn\speAF{
\omega(\mu+i0)+\omega(\mu-i0)=
-\zeta+{\rm sign}(\mu)\qquad\forall\mu\in[\a,\a']\cup[\b',\b]
}
with $\zeta=t/\gamma$.

We could proceed as in the previous sections; this
would lead to a representation of $\omega(z)$
in terms of elliptic integrals. However, this would be fairly cumbersome
and we proceed instead as follows. Introduce an elliptic parameterization
\eqn\ellparam{
u(\mu)={1\over2}\sqrt{(\b'-\a)(\b-\a')}\int_\b^\mu {\d z\over
\sqrt{(z-\a)(z-\a')(z-\b')(z-\b)}}
}
which corresponds to setting: 
${\b'-\a\over \b-\a}{\b-\mu\over \b'-\mu}=\sn^2(u,k)$ with
$k=\sqrt{(\b-\a)(\b'-\a')\over(\b'-\a)(\b-\a')}$. With an appropriate choice
of path of integration, this maps
the $\mu$ complex plane (resp.\ upper half-plane, lower half-plane)
onto the rectangle $[0,K]\times[-iK',iK']$
(resp.\ $[0,K]\times[0,iK']$, $[0,K]\times[-iK',0]$), where
$K$ and $K'$ are the usual complete elliptic integrals
of the first kind. Similarly,
the second sheet of the double covering is mapped onto the other
half of the torus, which can be chosen to be $[-K,0]\times[-iK',iK']$.
The point of this parameterization is that the resolvent $\omega$
is now a well-defined function of $u$. In fact we have the following 
properties:
\smallskip
\item{(i)} The function $\omega(u)$ can be extended to
a holomorphic function in the whole $u$ plane.
\smallskip
\item{(ii)} The function $\omega(u)$ satisfies the following functional
relations (for all complex $u$):
\eqna\funcrel
$$\eqalignno{
\omega(u+2iK')&=\omega(u)-{i\pi\over\gamma}&\funcrel{a}\cr
\omega(u+2K)&=\omega(u)-2&\funcrel{b}\cr
\omega(u)+\omega(-u)&=1-\zeta&\funcrel{c}\cr
}
$$

\noindent Eq.~\funcrel{a} is the analytic continuation of Eq.~\sat.
Similarly, by combining the analytic continuations
of the two equations contained in \speAF,
one obtains Eqs.~\funcrel{b,c}.
\smallskip
\item{(iii)} The function $\omega(u)$ has the following expansion
near $u_\infty=u(z=\infty)$:
\eqn\infexp{
\omega=-{2\over\sqrt{(\b'-\a)(\b-\a')}}
(u-u_\infty)+O(u-u_\infty)^2
}

\noindent This is a rewriting of the condition $\omega(z)\sim {1\over z}$ at infinity.
\smallskip
Using properties (i) and (ii) (Eqs.~\funcrel{a,b}),
we conclude that ${\d\over\d u}\omega(u)$
is a doubly periodic holomorphic function, and so is a constant.
In order to restore the coefficients of $\omega(u)$ we can use
properties (ii) or (iii). We find that
\eqn\omfin{
\omega(u)=-{1\over K}(u-u_\infty)
}
plus several conditions relating the different parameters of the problem:
\eqna\conds
$$\eqalignno{
{K'\over K}&={\pi\over 2\gamma}&\conds{a}\cr
\sqrt{(\b'-\a)(\b-\a')}&=2K&\conds{b}\cr
{u_\infty\over K}&={1-\zeta\over 2}&\conds{c}\cr
}
$$
Relation \conds{a} is particularly interesting since it shows that
the elliptic nome $q=\e{-\pi K'/K}=\e{-\pi^2/2\gamma}$
depends only on $\gamma$
(and not on $\zeta$). Also, the dual nome (under modular transformation)
$\tilde{q}=\e{-2\gamma}$ is up to a sign the
quantum group deformation parameter of the model.

We can rewrite the three conditions in terms of the endpoints; we find
\eqn\endpts{
\eqalign{
\b-\a&=2K {\dn u_\infty\over\sn u_\infty \cn u_\infty}\cr
\b-\a'&=2K {\cn u_\infty\over\sn u_\infty \dn u_\infty}\cr
\b-\b'&=2K {\cn u_\infty \dn u_\infty\over\sn u_\infty}\cr
}
}
In order to completely fix the four endpoints $\a$, $\a'$, $\b'$, $\b$, 
we need one extra relation; this is the equality of chemical potentials
in the two unsaturated regions. This relation takes the form
\eqn\chem{
\int_{\a'}^{\b'} (\omega(\mu+i0)+\omega(\mu-i0)) \d\mu
=(1-\zeta)\b'+(1+\zeta)\a'
}
Using the expression \omfin\ of $\omega(u)$, we can rewrite it as
\eqn\chemb{
\b'-(\b-\b'){\sn u_\infty\over\cn u_\infty \dn u_\infty} Z(u_\infty)=0
}
where $Z$ is Jacobi's Zeta function; this fixes $b'$ to be
\eqn\endptsb{
\b'=2K Z(u_\infty)
}
The endpoints are now determined by \endpts\ and \endptsb, supplemented
by the value \conds{c} of $u_\infty$.

At this point, we are ready to calculate the free energy. 
We first rewrite explicitly the resolvent (Eq.~\omfin) under the
form
\eqn\omfinb{
\omega(z)=
\int_z^\infty {\d z'\over\sqrt{(z'-\a)(z'-\a')(z'-\b')(z'-\b)}}
}
Next we expand it to order $1/z^2$ to find
\eqn\corAF{
{\der f\over\der\zeta}={\a+\a'+\b'+\b\over 4}
}
which generalizes Eq.~\cor; using some known identities satisfied
by Zeta and theta functions, we obtain
\eqn\corAFb{
{\der f\over\der\zeta}=-{\pi\over 2} 
{\theta'_2(\pi\zeta/2)\over\theta_2(\pi\zeta/2)}
}
where we recall that $\theta_2(z)$ is
\eqn\deftheta{
\theta_2(z)=2\sum_{n=0}^\infty
q^{(n+1/2)^2} \cos(2n+1)z
}
There are a variety of ways to find the integration constant.
One is to calculate explicitly $f$ (for a particular
value of $\zeta$, e.g.\ $\zeta=0$) using this matrix model
solution, and then restore the $\gamma$ dependence coming
from \discrAF; this is a straightforward but tedious exercise. Another
possibility is to use the known limits $\zeta\to\pm 1$, that is 
$t\to\pm\gamma$, where we should have (see \KZJ)
\eqn\lims{
\e{f}\sim{1\over \gamma \mp t}
}
Either way, we finally find:
\eqn\finalAF{
\e{f}={\pi\over 2\gamma}{\theta_1'(0)\over\theta_2({\pi t\over 2\gamma})}
}
where we recall that the elliptic nome is $q=\e{-{\pi^2\over 2\gamma}}$.

As a simple check of our calculation, note that if one
sends $\gamma$ to $0$ (keeping $\zeta$ fixed), 
since the constraint \constr, which
was the only difference with the disordered phase,
disappears, one should recover the results of the previous section.
This is indeed what happens when one replaces the theta functions
with their $q\to 0$ limit. Also, \finalAF\ has been 
numerically checked with high accuracy.

This concludes the calculation of the bulk free energy in
the anti-ferroelectric phase. 
Restated more explicitly, this is the result we have obtained:
the partition function $Z_N$ of the six-vertex
model on a $N\times N$ lattice with DWBC and 
Boltzmann weights given by \weiAF\ has the following large $N$ behavior:
\eqn\finalAFb{
\lim_{N\to\infty} Z_N^{1/N^2}=\sinh \gamma(1-\zeta)\sinh\gamma(1+\zeta)\,
{\pi\over 2\gamma}{\theta_1'(0)\over\theta_2({\pi \zeta\over 2})}
}
where $\zeta=t/\gamma$, and
the elliptic nome of the theta functions is $q=\e{-{\pi^2\over 2\gamma}}$.
Note that
this expression is different from the corresponding expression for PBC.
Let us now consider the two limits $\gamma\to 0$ and $\gamma\to\infty$.
In both cases we shall assume that $\zeta$ remains fixed.

\subsec{Small $\gamma$ limit}
As one sends $\gamma$ to $0$, one reaches the line of 
the disordered/anti-ferroelectric phase
transition. As noted earlier, the bulk free energy of the
disordered phase is essentially obtained
from that of the anti-ferroelectric phase by setting $q=0$ in the
theta functions (and performing the rotation $\gamma\to i\gamma$,
$t\to it$
in the prefactors). Considering that $q=\e{-{\pi^2\over 2\gamma}}$,
we expect a very smooth phase transition. More explicitly, we have
the following expansion of $f$:
\eqn\expAF{
f=\log \left[{\pi\over 2\gamma} {1\over\cos({\pi t\over 2\gamma})}\right]
+2\sum_{m=1}^\infty {1\over m}
{q^{2m}\over 1-q^{2m}} (1-(-1)^m \cos(m\pi t/\gamma))
}
After subtraction of the analytic continuation of the disordered phase free energy
(note that this analytic continuation is trivial since $f$ only depends on $t/\gamma$),
we obtain the singular part of the free energy, which has a leading singularity
\eqn\Fsing{
f_{\rm sing}=4\, \e{-\pi^2/\gamma} \cos^2\left({\pi t\over 2\gamma}\right)+\cdots
}
This is the same type of singularity that appears
in the model with periodic boundary conditions \Bax. In more physical
terms, if we introduce a temperature $T$ which is
near the critical temperature $T_c$, we have
\eqn\Fsingb{
f_{\rm sing} \propto \e{-C/\sqrt{T_c-T}}
}
that is an infinite order phase transition.

\subsec{Large $\gamma$ limit}
Next, let us consider the $\gamma\to\infty$ limit, i.e.\ 
$\Delta=-\cosh(2\gamma)\to -\infty$. This is a typical zero temperature
limit, and we expect that the free energy will be dominated by
the contribution of a ground state. After
a modular transformation, the bulk free energy reads
\eqn\modtransf{
\eqalign{
F&= -\log(\sinh(\gamma-t)\sinh(\gamma+t))-f\cr
&=
-{\gamma\over 2}-{t^2\over 2\gamma}
-\log\sinh(\gamma+t)+t
+2\sum_{m=1}^\infty {1\over m}{\e{-2m\gamma}\over\sinh(2m\gamma)}\sinh^2(m(\gamma-t))
\cr}
}
We can interpret the first terms when $\gamma\to\infty$
\eqn\leadlowT{
F= -{3\over2}\gamma-{t^2\over 2\gamma}+O(\e{-2\gamma})
}
as coming from the family of ground states described by Fig.~\afgs.
The pattern of a rectangle inscribed inside a square is reminiscent
of the circle inscribed inside a square characteristic
of the disordered phase \JPS.
\fig\afgs{Ground states of the anti-ferroelectric 
phase. In regions $a$ and $b$ the arrows are aligned,
whereas in region $c$ they alternate in 
direction.}{\epsfxsize=5cm
\epsfbox{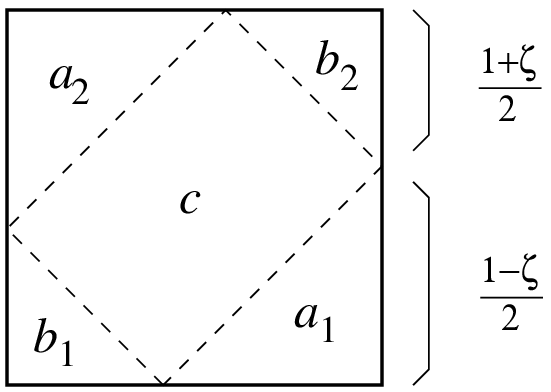}}

\subsec{Subdominant corrections}
As a final note, it is interesting to understand why
the approach of \KZJ\ fails in the anti-ferroelectric phase.
There, the idea was to find an appropriate Ansatz
on the asymptotic behavior of the determinant $\tau_N$ 
and plug it in
the Toda equation \hirota. The simplest assumption is that
only the leading behavior (bulk free energy) must be taken into
account, which leads to replacing
$\tau_N$ with $c_N\, \e{N^2 f}$, where
$c_N=(\prod_{n=0}^{N-1} n!)^2$.
The Toda equation then reduces to the ordinary differential
equation for $f$:
\eqn\diff{
f''=\e{2f}
}

We can now use some insight from matrix models to understand whether this
assumption was justified or not. In the ferroelectric and disordered
phases, we reduced the computation of $\tau_N$ to a matrix model
with eigenvalues in 
{\it one single interval} $[a,b]$ (disregarding the saturated region which
plays no role here); it is known that such models have a regular
large $N$ limit. In fact, in the ferroelectric phase one
can easily prove that
\eqn\smoothasyb{
\tau_N \sim c_N\, \e{N^2 f}\,\e{N(t-|\gamma|)}
}
up to only exponentially small corrections; whereas in the disordered phase,
one expects an asymptotic expansion which starts with
\eqn\smoothasy{
\tau_N \sim c_N\,
\e{N^2 f}\, N^{\kappa}\, C
}
and continues with inverse powers of $N$ 
(note that this is not quite the usual topological expansion of $2D$ gravity
since the potential is not polynomial).
In either case, the
assumption on the corrections
is valid, and indeed, one can check that the expressions
\finalF\ and \finalD\ do satisfy the ODE \diff.

On the contrary, in the anti-ferroelectric regime, we have found
that the support of the eigenvalues
contains {\it two intervals} $[a,a']$ and $[b',b]$ and therefore
we expect to be in a situation similar to what was studied
in \refs{\Deift,\BDE}. The analysis shows that $\tau_N$ should in
this case display a pseudo-periodic behavior, which is indeed
what is found in numerical computations. More precisely, after some
calculations along the lines of \BDE, one finds that
\eqn\notsmoothasy{
\tau_N \sim c_N\,
\left[{\pi\over 2\gamma}{\theta_1'(0)\over\theta_2({\pi \zeta\over 2})}
\right]^{N^2}
\theta_4\left({\pi\over2}(1+\zeta)N\right)
\,C
}
where $\zeta=t/\gamma$ and
the elliptic nome $q$ of the theta function is as before
$q=\e{-{\pi^2\over 2\gamma}}$. The constant $C$ depends only on $\gamma$. 
One can check that the right hand side of Eq.~\notsmoothasy\ 
does satisfy the Toda equation
\hirota, even though the
bulk free energy alone does not satisfy the ODE \diff.
%With a bit more work, the constant $C(\gamma)$ can be determined: (?)
\bigskip
\centerline{\bf Acknowledgements}
I thank V.~Korepin for discussions, and the Centre de Recherches
Math\'ematiques de l'Universit\'e de Montr\'eal,
where part of this work was performed, for its hospitality.
\footatend\vfill\supereject\immediate\closeout\rfile\writestoppt
\baselineskip=14pt\centerline{{\bf References}}\bigskip{\frenchspacing%
\parindent=20pt\escapechar=` \input refs.tmp\vfill\eject}\nonfrenchspacing
\bye

%% file: mssymb.tex
%               *****     MSSYMB.TeX    *****                  8 Jul 87

%

%       This file contains the definitions for the symbols in the two

%       "extra symbols" fonts created at the American Math. Society.

%

%       Ce fichier a 't' modifi' le 20 Septembre 1991, pour remplacer

%       les anciennes fontes msxm et msym par les fontes plus r'centes

%       msam et msbm. Le nom du fichier n'a pas 't' chang'.

\catcode`\@=11

\font\tenmsa=msam10

\font\sevenmsa=msam7

\font\fivemsa=msam5

\font\tenmsb=msbm10

\font\sevenmsb=msbm7

\font\fivemsb=msbm5

\newfam\msafam

\newfam\msbfam

\textfont\msafam=\tenmsa  \scriptfont\msafam=\sevenmsa

  \scriptscriptfont\msafam=\fivemsa

\textfont\msbfam=\tenmsb  \scriptfont\msbfam=\sevenmsb

  \scriptscriptfont\msbfam=\fivemsb

\def\hexnumber@#1{\ifcase#1 0\or1\or2\or3\or4\or5\or6\or7\or8\or9\or

	A\or B\or C\or D\or E\or F\fi }

%  The following 13 lines establish the use of the Euler Fraktur font.

%  To use this font, remove % from beginning of these lines.

\font\teneuf=eufm10

\font\seveneuf=eufm7

\font\fiveeuf=eufm5

\newfam\euffam

\textfont\euffam=\teneuf

\scriptfont\euffam=\seveneuf

\scriptscriptfont\euffam=\fiveeuf

\def\frak{\ifmmode\let\next\frak@\else

 \def\next{\Err@{Use \string\frak\space only in math mode}}\fi\next}

\def\goth{\ifmmode\let\next\frak@\else

 \def\next{\Err@{Use \string\goth\space only in math mode}}\fi\next}

\def\frak@#1{{\frak@@{#1}}}

\def\frak@@#1{\fam\euffam#1}

%  End definition of Euler Fraktur font.

\edef\msa@{\hexnumber@\msafam}

\edef\msb@{\hexnumber@\msbfam}

\mathchardef\boxdot="2\msa@00

\mathchardef\boxplus="2\msa@01

\mathchardef\boxtimes="2\msa@02

\mathchardef\square="0\msa@03

\mathchardef\blacksquare="0\msa@04

\mathchardef\centerdot="2\msa@05

\mathchardef\lozenge="0\msa@06

\mathchardef\blacklozenge="0\msa@07

\mathchardef\circlearrowright="3\msa@08

\mathchardef\circlearrowleft="3\msa@09

\mathchardef\rightleftharpoons="3\msa@0A

\mathchardef\leftrightharpoons="3\msa@0B

\mathchardef\boxminus="2\msa@0C

\mathchardef\Vdash="3\msa@0D

\mathchardef\Vvdash="3\msa@0E

\mathchardef\vDash="3\msa@0F

\mathchardef\twoheadrightarrow="3\msa@10

\mathchardef\twoheadleftarrow="3\msa@11

\mathchardef\leftleftarrows="3\msa@12

\mathchardef\rightrightarrows="3\msa@13

\mathchardef\upuparrows="3\msa@14

\mathchardef\downdownarrows="3\msa@15

\mathchardef\upharpoonright="3\msa@16

\mathchardef\downharpoonright="3\msa@17

\mathchardef\upharpoonleft="3\msa@18

\mathchardef\downharpoonleft="3\msa@19

\mathchardef\rightarrowtail="3\msa@1A

\mathchardef\leftarrowtail="3\msa@1B

\mathchardef\leftrightarrows="3\msa@1C

\mathchardef\rightleftarrows="3\msa@1D

\mathchardef\Lsh="3\msa@1E

\mathchardef\Rsh="3\msa@1F

\mathchardef\rightsquigarrow="3\msa@20

\mathchardef\leftrightsquigarrow="3\msa@21

\mathchardef\looparrowleft="3\msa@22

\mathchardef\looparrowright="3\msa@23

\mathchardef\circeq="3\msa@24

\mathchardef\succsim="3\msa@25

\mathchardef\gtrsim="3\msa@26

\mathchardef\gtrapprox="3\msa@27

\mathchardef\multimap="3\msa@28

\mathchardef\therefore="3\msa@29

\mathchardef\because="3\msa@2A

\mathchardef\doteqdot="3\msa@2B

\mathchardef\triangleq="3\msa@2C

\mathchardef\precsim="3\msa@2D

\mathchardef\lesssim="3\msa@2E

\mathchardef\lessapprox="3\msa@2F

\mathchardef\eqslantless="3\msa@30

\mathchardef\eqslantgtr="3\msa@31

\mathchardef\curlyeqprec="3\msa@32

\mathchardef\curlyeqsucc="3\msa@33

\mathchardef\preccurlyeq="3\msa@34

\mathchardef\leqq="3\msa@35

\mathchardef\leqslant="3\msa@36

\mathchardef\lessgtr="3\msa@37

\mathchardef\backprime="0\msa@38

\mathchardef\risingdotseq="3\msa@3A

\mathchardef\fallingdotseq="3\msa@3B

\mathchardef\succcurlyeq="3\msa@3C

\mathchardef\geqq="3\msa@3D

\mathchardef\geqslant="3\msa@3E

\mathchardef\gtrless="3\msa@3F

\mathchardef\sqsubset="3\msa@40

\mathchardef\sqsupset="3\msa@41

\mathchardef\vartriangleright="3\msa@42

\mathchardef\vartriangleleft="3\msa@43

\mathchardef\trianglerighteq="3\msa@44

\mathchardef\trianglelefteq="3\msa@45

\mathchardef\bigstar="0\msa@46

\mathchardef\between="3\msa@47

\mathchardef\blacktriangledown="0\msa@48

\mathchardef\blacktriangleright="3\msa@49

\mathchardef\blacktriangleleft="3\msa@4A

\mathchardef\vartriangle="0\msa@4D

\mathchardef\blacktriangle="0\msa@4E

\mathchardef\triangledown="0\msa@4F

\mathchardef\eqcirc="3\msa@50

\mathchardef\lesseqgtr="3\msa@51

\mathchardef\gtreqless="3\msa@52

\mathchardef\lesseqqgtr="3\msa@53

\mathchardef\gtreqqless="3\msa@54

\mathchardef\Rrightarrow="3\msa@56

\mathchardef\Lleftarrow="3\msa@57

\mathchardef\veebar="2\msa@59

\mathchardef\barwedge="2\msa@5A

\mathchardef\doublebarwedge="2\msa@5B

\mathchardef\angle="0\msa@5C

\mathchardef\measuredangle="0\msa@5D

\mathchardef\sphericalangle="0\msa@5E

\mathchardef\varpropto="3\msa@5F

\mathchardef\smallsmile="3\msa@60

\mathchardef\smallfrown="3\msa@61

\mathchardef\Subset="3\msa@62

\mathchardef\Supset="3\msa@63

\mathchardef\Cup="2\msa@64

\mathchardef\Cap="2\msa@65

\mathchardef\curlywedge="2\msa@66

\mathchardef\curlyvee="2\msa@67

\mathchardef\leftthreetimes="2\msa@68

\mathchardef\rightthreetimes="2\msa@69

\mathchardef\subseteqq="3\msa@6A

\mathchardef\supseteqq="3\msa@6B

\mathchardef\bumpeq="3\msa@6C

\mathchardef\Bumpeq="3\msa@6D

\mathchardef\lll="3\msa@6E

\mathchardef\ggg="3\msa@6F

\mathchardef\circledS="0\msa@73

\mathchardef\pitchfork="3\msa@74

\mathchardef\dotplus="2\msa@75

\mathchardef\backsim="3\msa@76

\mathchardef\backsimeq="3\msa@77

\mathchardef\complement="0\msa@7B

\mathchardef\intercal="2\msa@7C

\mathchardef\circledcirc="2\msa@7D

\mathchardef\circledast="2\msa@7E

\mathchardef\circleddash="2\msa@7F

\def\ulcorner{\delimiter"4\msa@70\msa@70 }

\def\urcorner{\delimiter"5\msa@71\msa@71 }

\def\llcorner{\delimiter"4\msa@78\msa@78 }

\def\lrcorner{\delimiter"5\msa@79\msa@79 }

\def\yen{\mathhexbox\msa@55 }

\def\checkmark{\mathhexbox\msa@58 }

\def\circledR{\mathhexbox\msa@72 }

\def\maltese{\mathhexbox\msa@7A }

\mathchardef\lvertneqq="3\msb@00

\mathchardef\gvertneqq="3\msb@01

\mathchardef\nleq="3\msb@02

\mathchardef\ngeq="3\msb@03

\mathchardef\nless="3\msb@04

\mathchardef\ngtr="3\msb@05

\mathchardef\nprec="3\msb@06

\mathchardef\nsucc="3\msb@07

\mathchardef\lneqq="3\msb@08

\mathchardef\gneqq="3\msb@09

\mathchardef\nleqslant="3\msb@0A

\mathchardef\ngeqslant="3\msb@0B

\mathchardef\lneq="3\msb@0C

\mathchardef\gneq="3\msb@0D

\mathchardef\npreceq="3\msb@0E

\mathchardef\nsucceq="3\msb@0F

\mathchardef\precnsim="3\msb@10

\mathchardef\succnsim="3\msb@11

\mathchardef\lnsim="3\msb@12

\mathchardef\gnsim="3\msb@13

\mathchardef\nleqq="3\msb@14

\mathchardef\ngeqq="3\msb@15

\mathchardef\precneqq="3\msb@16

\mathchardef\succneqq="3\msb@17

\mathchardef\precnapprox="3\msb@18

\mathchardef\succnapprox="3\msb@19

\mathchardef\lnapprox="3\msb@1A

\mathchardef\gnapprox="3\msb@1B

\mathchardef\nsim="3\msb@1C

%\mathchardef\napprox="3\msb@1D

\mathchardef\ncong="3\msb@1D

\mathchardef\varsubsetneq="3\msb@20

\mathchardef\varsupsetneq="3\msb@21

\mathchardef\nsubseteqq="3\msb@22

\mathchardef\nsupseteqq="3\msb@23

\mathchardef\subsetneqq="3\msb@24

\mathchardef\supsetneqq="3\msb@25

\mathchardef\varsubsetneqq="3\msb@26

\mathchardef\varsupsetneqq="3\msb@27

\mathchardef\subsetneq="3\msb@28

\mathchardef\supsetneq="3\msb@29

\mathchardef\nsubseteq="3\msb@2A

\mathchardef\nsupseteq="3\msb@2B

\mathchardef\nparallel="3\msb@2C

\mathchardef\nmid="3\msb@2D

\mathchardef\nshortmid="3\msb@2E

\mathchardef\nshortparallel="3\msb@2F

\mathchardef\nvdash="3\msb@30

\mathchardef\nVdash="3\msb@31

\mathchardef\nvDash="3\msb@32

\mathchardef\nVDash="3\msb@33

\mathchardef\ntrianglerighteq="3\msb@34

\mathchardef\ntrianglelefteq="3\msb@35

\mathchardef\ntriangleleft="3\msb@36

\mathchardef\ntriangleright="3\msb@37

\mathchardef\nleftarrow="3\msb@38

\mathchardef\nrightarrow="3\msb@39

\mathchardef\nLeftarrow="3\msb@3A

\mathchardef\nRightarrow="3\msb@3B

\mathchardef\nLeftrightarrow="3\msb@3C

\mathchardef\nleftrightarrow="3\msb@3D

\mathchardef\divideontimes="2\msb@3E

\mathchardef\varnothing="0\msb@3F

\mathchardef\nexists="0\msb@40

\mathchardef\mho="0\msb@66

\mathchardef\eth="0\msb@67

\mathchardef\eqsim="3\msb@68

\mathchardef\beth="0\msb@69

\mathchardef\gimel="0\msb@6A

\mathchardef\daleth="0\msb@6B

\mathchardef\lessdot="3\msb@6C

\mathchardef\gtrdot="3\msb@6D

\mathchardef\ltimes="2\msb@6E

\mathchardef\rtimes="2\msb@6F

\mathchardef\shortmid="3\msb@70

\mathchardef\shortparallel="3\msb@71

\mathchardef\smallsetminus="2\msb@72

\mathchardef\thicksim="3\msb@73

\mathchardef\thickapprox="3\msb@74

\mathchardef\approxeq="3\msb@75

\mathchardef\succapprox="3\msb@76

\mathchardef\precapprox="3\msb@77

\mathchardef\curvearrowleft="3\msb@78

\mathchardef\curvearrowright="3\msb@79

\mathchardef\digamma="0\msb@7A

\mathchardef\varkappa="0\msb@7B

\mathchardef\hslash="0\msb@7D

\mathchardef\hbar="0\msb@7E

\mathchardef\backepsilon="3\msb@7F

\def\Bbb{\ifmmode\let\next\Bbb@\else

 \def\next{\errmessage{Use \string\Bbb\space only in math mode}}\fi\next}

\def\Bbb@#1{{\Bbb@@{#1}}}

\def\Bbb@@#1{\fam\msbfam#1}

\catcode`\@=12